\documentclass[11pt]{article}
\usepackage{amsmath}
\usepackage{amsfonts}
\usepackage[dvips]{epsfig}

 \advance \textwidth by -2in
 \advance \textheight by -3in
\def\##1{{\bf #1}}
\def\=#1{\underline{\underline{#1}}}

\def\+#1{\underline{\bf #1}}
\def\*#1{\underline{\underline{\bf #1}}}

\def\eps{\epsilon}
\def\epso{\epsilon_{\scriptscriptstyle 0}}
\def\muo{\mu_{\scriptscriptstyle 0}}
\def\ko{k_{\scriptscriptstyle 0}}
\def\wo{w_{\scriptscriptstyle 0}}

\def\lambdao{\lambda_{\scriptscriptstyle 0}}
\def\etao{\eta_{\scriptscriptstyle 0}}
\def\.{\mbox{ \tiny{$^\bullet$} }}

\def\le{\left(}
\def\ri{\right)}
\def\les{\left[}
\def\ris{\right]}
\def\lec{\left\{}
\def\ric{\right\}}

\def\l#1{\label{#1}}
\def\r#1{(\ref{#1})}

\begin{document}

\noindent \LARGE{{\bf Concealment by uniform motion }} \vskip 0.2cm

\normalsize

\noindent  {\bf Tom G. Mackay$^1$ and Akhlesh Lakhtakia$^2$} \vskip
0.2cm

\noindent {\sf $^1$ School of Mathematics\\
\noindent James Clerk Maxwell Building\\
\noindent University of Edinburgh\\
\noindent Edinburgh EH9 3JZ, UK \\ \noindent email:
T.Mackay@ed.ac.uk}
 \vskip 0.4cm

\noindent {\sf $^2$ CATMAS~---~Computational \& Theoretical Materials Sciences Group \\
\noindent Department of Engineering Science \& Mechanics\\
\noindent 212 Earth \& Engineering Sciences Building\\
\noindent Pennsylvania State University\\ \noindent University Park,
PA, 16802--6812, USA \\ \noindent email: akhlesh@psu.edu } \vskip
0.4cm

\

\begin{abstract}
The perceived lateral position of a transmitted beam, upon
propagating through
 a slab made of homogeneous, isotropic, dielectric material at an oblique angle,
  can be controlled through varying the velocity of the slab.
In particular, by judiciously selecting the slab velocity, the
transmitted beam can emerge from the slab with no lateral shift in
position. Thereby, a degree of concealment can be achieved. This
concealment is explored in numerical calculations based on a 2D
Gaussian beam.
\end{abstract}

 \vskip
0.4cm


\section{Introduction}

The topic of invisibility~---~which is a very old one in optics and
electromagnetics (Wolf \& Habashy 1993)~---~has lately acquired a
new lease of life with the advent of  metamaterials (Fedotov,
Mladyonov, Prosvirnin \& Zheludev 2005; Al\`{u} \& Engheta 2006). In
a similar vein,
 it has recently  been proposed that the exotic electromagnetic
possibilities offered by metamaterials may be harnessed to achieve
cloaking, at least in principle (Leonhardt 2006; Milton \&
Nicorovici 2006; Pendry, Schurig \& Smith 2006). The  theoretical
arguments underlying this proposed cloaking are based on the
facilitation of coordinate transformations by nonhomogeneous
metamaterials.

 A quite different approach to concealing a material
  is pursued in this paper. It is based on the perceived deflection of light by a
material slab translating at constant velocity. No special material properties
are required, but  for illustrative purposes, we consider an isotropic,
homogeneous, dielectric material. In  two previous studies, we have
demonstrated that the much--heralded negative--phase--velocity
phenomenon often associated with
negatively refracting electromagnetic metamaterials can be realized by
conventional materials through the process of uniform motion (Mackay
\& Lakhtakia 2004, 2006a). Here, we demonstrate that a substantial
degree of concealment may also be realized  by uniform motion.

 As regards notational matters,  3 vectors are in boldface, with
the $\hat{}$ symbol identifying  unit vectors. Double underlining
signifies a 3$\times$3 dyadic and $\=I$ is the identity 3$\times$3
dyadic. The superscript ${}^T$ denotes the transpose of a column
vector.
The permittivity and permeability of vacuum are  $\epso$ and
$\muo$. The
 vacuum wavenumber is $\ko = \omega \sqrt{\epso \muo}$ with
 $\omega$ being the angular frequency, and the vacuum wavelength is $\lambdao = 2 \pi / \ko$.

\section{Planewave propagation into a uniformly moving half--space}
\l{pw_section}

\begin{figure}[!ht]
\centering \psfull \epsfig{file=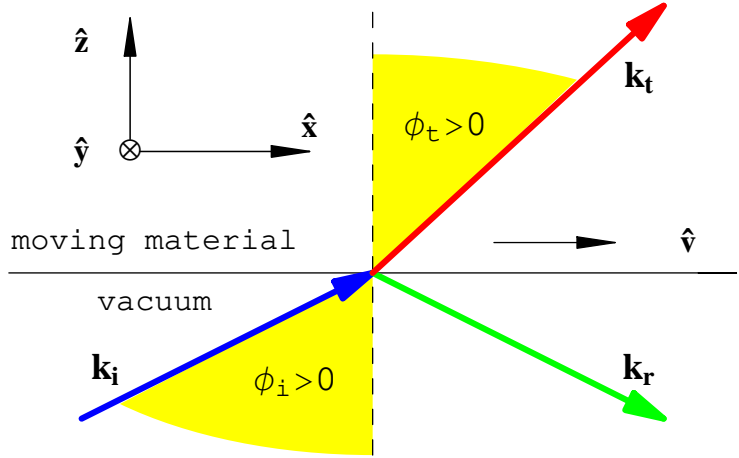,width=4.0in}
 \caption{\label{fig1}
 A plane wave with wavevector $\#k_i$
  is incident from vacuum onto a half--space occupied by a simply moving
  dielectric material at an angle  $\phi_i$ with
respect to the unit vector $\hat{\#z}$ normal to the planar
interface. The moving material is characterized by relative
permittivity $\eps_r > 0$  in a co--moving frame of reference. As
observed in the non--co--moving (laboratory) frame of reference
wherein the incident plane wave is specified, the refracted
wavevector $\#k_t$ makes an angle $\phi_t$ with $\hat{\#z}$.
  }
\end{figure}

As a preliminary to concealment of a moving slab
(\S\ref{Beam_section}), let us consider a uniformly moving
half--space.
 Suppose that a plane wave is launched with wavevector
$\#k_i = k_i \hat{\#k}_i$ from vacuum ($z<0$) towards the half--space
$z>0$ occupied
by an isotropic, nondissipative, dielectric material. This material
moves at constant velocity $\#v = v \hat{\#v} = v \hat{\#x}$,
parallel to the interface and in the plane of incidence. In an inertial frame of reference that moves with the same
velocity $\#v$ with respect to the laboratory frame of reference
wherein $\#k_i$ is specified, the refracting material is characterized
by relative permittivity $\eps_r $.
The Minkowski
constitutive relations of the moving half--space in the laboratory
frame of reference are (Chen 1983)
\begin{equation}
\left.
\begin{array}{l}
\#D(\#r) = \epso \eps_r \, \=\alpha \. \#E(\#r) + \sqrt{\epso \muo}
\le \#m \times \=I \,\ri \. \#H(\#r) \vspace{4pt} \\
\#B(\#r) = - \sqrt{\epso \muo} \le \#m \times \=I \,\ri \. \#E(\#r)
+ \muo \, \=\alpha \. \#H(\#r)
\end{array}\right\}, \l{CRs}
\end{equation}
where
\begin{equation}
\left.
\begin{array}{lll}
&\=\alpha = \alpha \, \=I + \le 1 - \alpha \ri\, \hat{\#v}
\hat{\#v}\,, & \displaystyle{\alpha = \frac{1-\beta^2}{1 - \eps_r
\beta^2}}\,,
\\[5pt]
&\#m = m \hat{\#v}\,,  \quad \displaystyle{ m= \frac{\beta \le
\eps_r - 1 \ri}{1 - \eps_r \beta^2}}\,, &\beta = v \sqrt{\epso
\muo}\,
\end{array}
\right\}\,.
\end{equation}
In order to exclude the
possibility of evanescent plane waves, $\eps_r > 1$ is assumed. The
envisaged scenario is  illustrated schematically in
Figure~\ref{fig1}.

The angle $\phi_t$ between the refracted wavevector $\#k_t = k_t
\hat{\#k}_t$, as observed from the laboratory frame of reference,
and the unit vector $\hat{\#z}$ normal to the interface is related
to the angle of incidence
$
\phi_i = \cos^{-1} \le \hat{\#k}_i \.\hat{\#z}
   \ri
$
 by (Chen 1983)
\begin{equation}
  \phi_t = \sin^{-1}  \le \frac{\ko \sin
\phi_i}{k_t} \ri \,,
\end{equation}
where
\begin{equation}
  k_t =
  \ko \lec 1 + \xi \les 1 -
\beta \le \hat{\#k}_i \. \hat{\#v} \ri \ris^2 \ric^{1/2}
\end{equation}
is the wavenumber of the refracted wave and
$
   \xi =
{(\eps_r  - 1)}/{(1 - \beta^2)}$.
Since $0 < \phi_t <
\pi/2\,\forall\,\phi_i \in (0, \pi/2)$, refraction is positive $\forall\,\beta
\in (-1,1)$ (Mackay \& Lakhtakia 2006b).

\begin{figure}[!hb]
\centering \psfull \epsfig{file=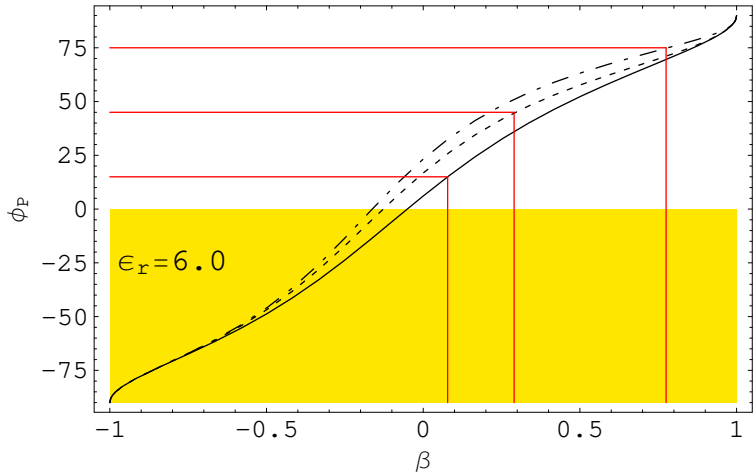,width=4.0in}
 \caption{\label{fig2} The angle $\phi_P$ (in degree)
 between the time--averaged Poynting vector $\#P_t$ and
  the unit vector $\hat{\#z}$,
   plotted as a function of $\beta \in (-1,1)$, when the angle of
incidence $\phi_i = 15^\circ$ (solid curve), $45^\circ$ (dashed
curve) and $75^\circ$ (broken dashed curve); and  $\eps_r = 6.0$.
The red  lines indicate where $\phi_P = \phi_i$.  The
counterposition regime  $\left\{\phi_P < 0^\circ, \phi_t>0^\circ\right\}$ is shaded.
  }
\end{figure}

The time--averaged Poynting vector of the refracted plane wave is
given by (Chen 1983)
\begin{equation}
\#P_t = P_t \, \hat{\#P}_t = \le
 \,| C_1 |^2 +  \eps_r  | C_2 |^2 \ri \le
\#k_t \times \hat{\#v} \ri^2 \les \#k_t + \xi \beta \le \ko - \beta
\#k_t \. \hat{\#v} \ri \hat{\#v} \ris,
\end{equation}
where $C_{1}$ and $C_2$ are constants, and the angle between $\hat{\#z}$
and $\hat{\#P}_t$ is
\begin{equation}
 \phi_P = \tan^{-1} \le \frac{\hat{\#P}_t \.\#v}{
| v | \, \hat{\#P}_t \.\hat{\#z}}
   \ri.
\end{equation}
As an illustrative example,
 the angle $\phi_P$
  is plotted in Figure~\ref{fig2} against $\beta \in (-1,1)$ for
$\phi_i \in \lec 15^\circ, 45^\circ, 75^\circ \ric$,  for the
half--space characterized by $\eps_r = 6.0$. The orientation of the
refracted time--averaged Poynting vector rotates towards the
direction of motion as $\beta$ increases from $-1$. The
counterposition regime~---~which occurs where $\phi_P < 0$ for
$\phi_t > 0$~---~is discussed elsewhere (Lakhtakia \& McCall 2004;
Mackay \& Lakhtakia 2006b).

In connection with Figure~\ref{fig2}, it is of particular interest
here that $\phi_P = \phi_i$ at (i) $\beta = 0.08$ for $\phi_i = 15
^\circ$, (ii) $\beta = 0.29$ for $\phi_i = 45 ^\circ$, and (iii) $\beta = 0.78$
for $\phi_i = 75 ^\circ$. That is, there exist angles of incidence
at which the time--averaged Poynting vector is not deflected by the
uniformly moving half--space. This suggests that it may be possible
for a light  beam~---~not to be confused with a plane wave~---~to pass through a uniformly moving slab at an oblique angle
without experiencing a lateral shift in position. That suggestion inspired
the research presented in the next section.

\section{Beam propagation through a uniformly moving slab}
\l{Beam_section}

\begin{figure}[!ht]
\centering \psfull \epsfig{file=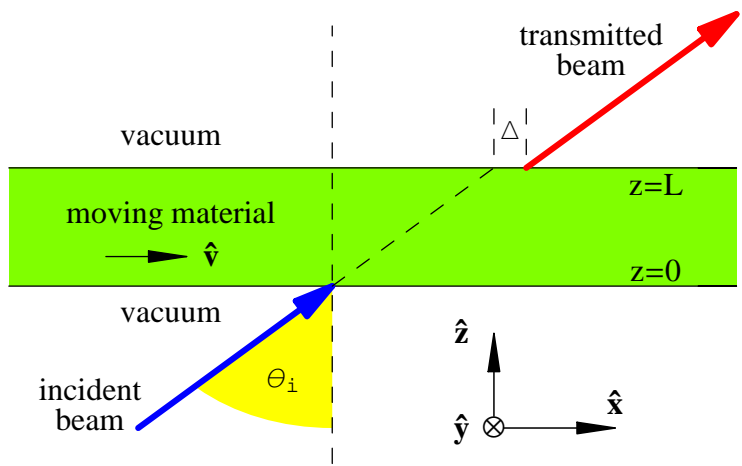,width=4.0in}
 \caption{\label{fig3} A beam is incident onto a simply moving slab at an angle  $\theta_i$ with
respect to the unit vector $\hat{\#z}$ normal to the planar
interface. The moving material is characterized by relative
permittivity $\eps_r > 0$  in a comoving frame of reference. As
observed in the non--comoving (laboratory) frame of reference
wherein the incident plane wave is specified, the transmitted beam
is shifted by $\Delta$, parallel to  $\hat{\#x}$,  relative to its
position if the slab were absent.
  }
\end{figure}

Suppose  that the uniformly moving half--space considered in
\S\ref{pw_section} is now replaced by a slab of thickness $L$ moving
at constant velocity ${\#v}=v\hat{{\#x}}$ parallel to its two surfaces,
as schematically illustrated in Figure~\ref{fig3}. The
slab~---~which is characterized, as before, by   $\eps_r > 1$
in a co--moving reference frame~---~is immersed in vacuum.

A 2D  beam with electric field  phasor (Haus 1984)
\begin{equation}
\#E_i \le x, z \ri = \int^{\infty}_{-\infty}
  \#e_i (\vartheta) \, \Psi (\vartheta)   \, \exp \les i
  \le \#k_i \. \#r  \ri
 \ris \; d \vartheta , \hspace{25mm} z \leq 0\l{Ei2}
\end{equation}
is incident upon the slab at a mean angle $\theta_i$ relative to the
slab normal direction $\mbox{\boldmath$\hat{z}$}$. The beam is
represented as an angular spectrum of plane waves, with
\begin{equation}
\#k_i = \ko \les \le \vartheta \, \cos \theta_i +
\sqrt{1-\vartheta^2} \, \sin \theta_i \ri \hat{\#x} - \le \vartheta
\, \sin \theta_i - \sqrt{1-\vartheta^2} \, \cos \theta_i \ri
\hat{\#z} \, \ris
\end{equation}
 being the wavevector
of each planewave contributor.
 The angular--spectral function $\Psi (\vartheta)$ is taken to have
 the Gaussian form (Haus 1984)
\begin{equation}
\Psi (\vartheta) = \frac{\ko \, \wo}{\sqrt{2 \pi}} \, \exp \les
-\frac{\le \ko \, \wo \, \vartheta \ri^2}{2}\ris,
\end{equation}
with $\wo$ being the width of the beam waist. Two polarization states
are considered: parallel to the plane of incidence, i.e.,
\begin{equation}
\#e_i (\vartheta) \equiv \#e_\parallel  =  \le \vartheta \, \sin
\theta_i - \sqrt{1-\vartheta^2} \, \cos \theta_i \ri \hat{\#x} + \le
\vartheta \, \cos \theta_i + \sqrt{1-\vartheta^2} \, \sin \theta_i
\ri \hat{\#z}
\end{equation}
 and perpendicular to the plane of incidence, i.e.,
\begin{equation}
\#e_i (\vartheta) \equiv \#e_\perp  = \hat{\#y}.
\end{equation}

The electric field  phasor of the reflected  beam is given as
\begin{equation}
\#E_r \le x, z \ri = \int^{\infty}_{-\infty}
  \#e_r (\vartheta) \, \Psi (\vartheta)   \, \exp \les i
  \le \#k_r \. \#r  \ri
 \ris \; d \vartheta, \hspace{25mm} z \leq 0, \l{Er}
\end{equation}
with
\begin{equation} \#k_r = \ko \les \le \vartheta \, \cos
\theta_i + \sqrt{1-\vartheta^2} \, \sin \theta_i \ri \hat{\#x} + \le
\vartheta \, \sin \theta_i - \sqrt{1-\vartheta^2} \, \cos \theta_i
\ri \hat{\#z} \, \ris
\end{equation}
and
\begin{equation}
\#e_r (\vartheta) = \left\{
\begin{array}{ccr}
r_\parallel \Big[ - \le \vartheta \, \sin \theta_i -
\sqrt{1-\vartheta^2} \, \cos \theta_i \ri \hat{\#x}&& \\ + \le
\vartheta \, \cos \theta_i + \sqrt{1-\vartheta^2} \, \sin \theta_i
\ri \hat{\#z} \Big] & \mbox{for} & \#e_i (\vartheta) = \#e_\parallel
\vspace{4pt}
 \\
r_\perp \,\#e_\perp & \mbox{for} &  \#e_i (\vartheta) = \#e_\perp
\end{array}
\right. .
\end{equation}
The electric field  phasor of the transmitted  beam is given as
\begin{equation}
\#E_t \le x, z \ri = \int^{\infty}_{-\infty}
  \#e_t (\vartheta) \, \Psi (\vartheta)   \, \exp \lec i
  \les \#k_t \. \le \#r - L \hat{\#z}   \ri \ris
 \ric \; d \vartheta, \hspace{25mm} z \geq L, \l{Et}
\end{equation}
with $\#k_t = \#k_i$ and
\begin{equation}
\#e_t (\vartheta) = \left\{
\begin{array}{ccr}
t_\parallel \, \#e_i (\vartheta) & \mbox{for} & \#e_i (\vartheta) =
\#e_\parallel \vspace{4pt}
 \\
t_\perp \, \#e_\perp  & \mbox{for} &  \#e_i (\vartheta) = \#e_\perp
\end{array}
\right. .
\end{equation}
Expressions for the  reflection coefficients $r_{\parallel, \perp}$
and transmission coefficients $t_{\parallel, \perp}$  are provided
in equations \r{r_perp}--\r{t_perp} in the Appendix.

\begin{figure}[!ht]
\centering \psfull \epsfig{file=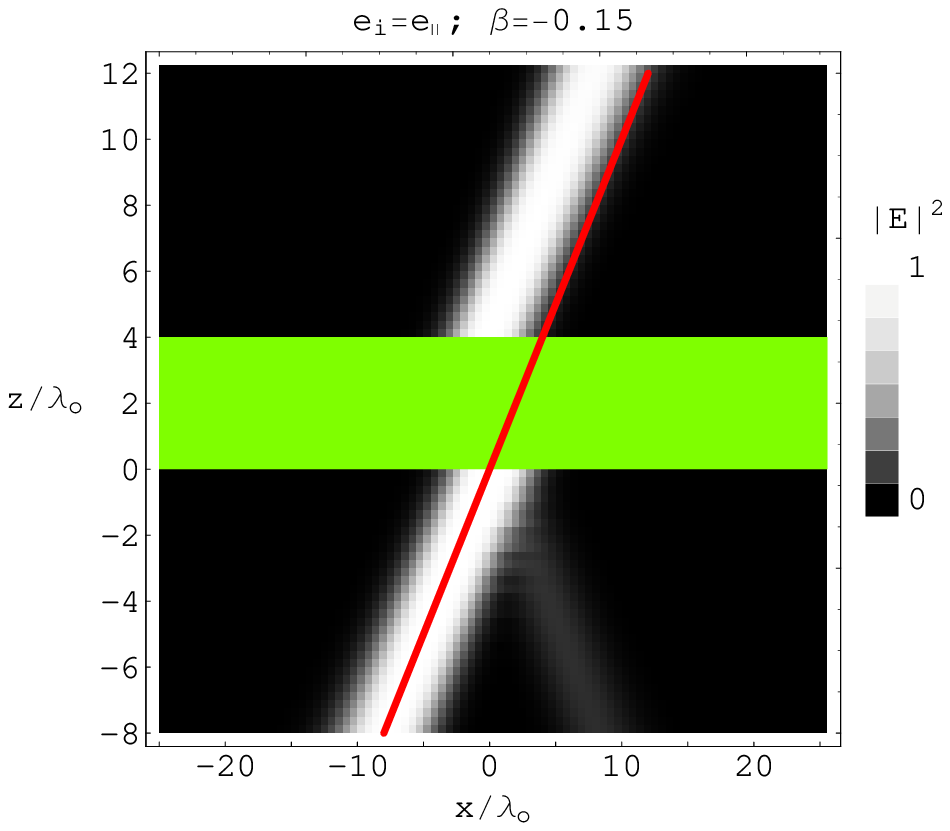,width=2.9in} \hfill
\epsfig{file=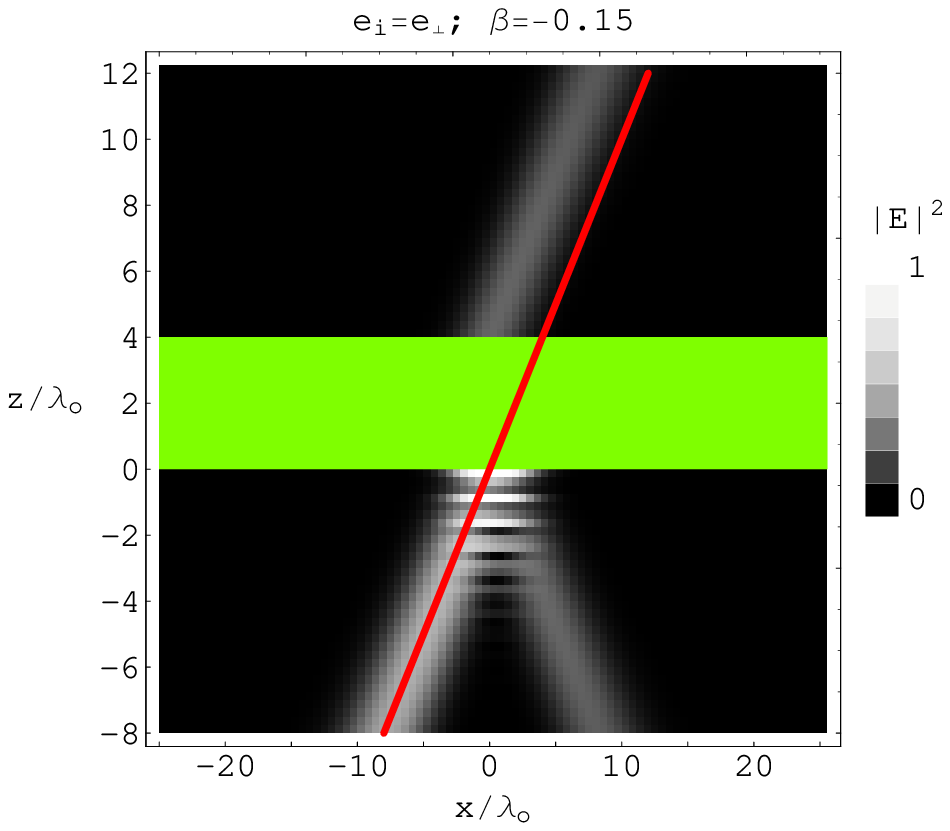,width=2.9in}\\
 \epsfig{file=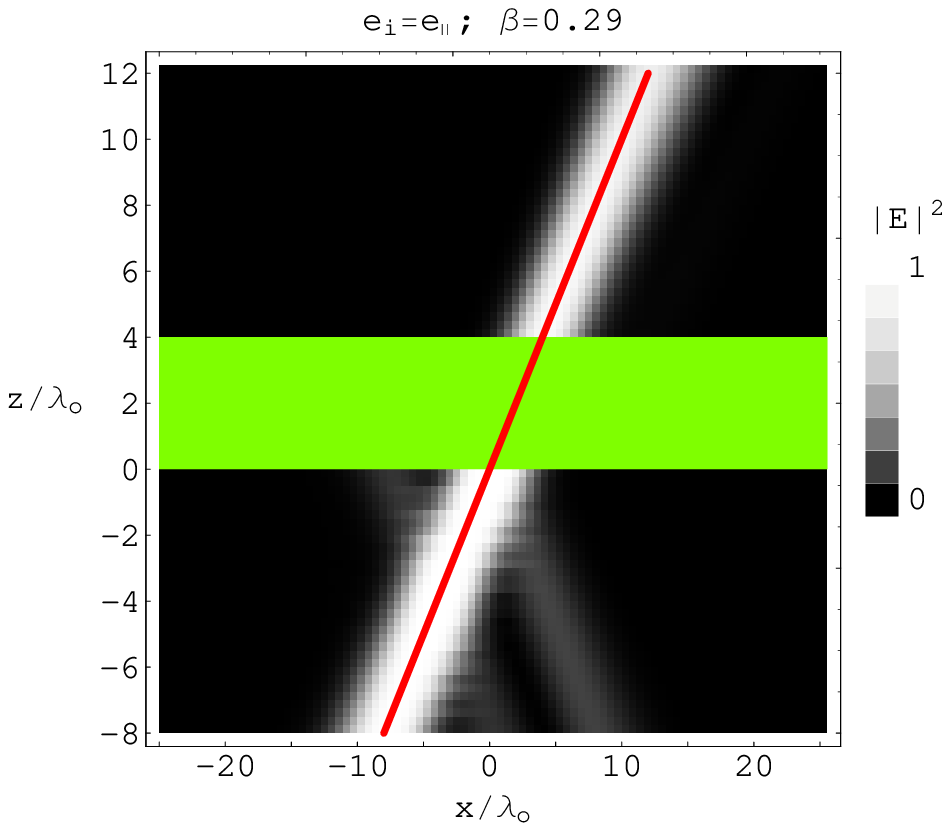,width=2.9in} \hfill
 \epsfig{file=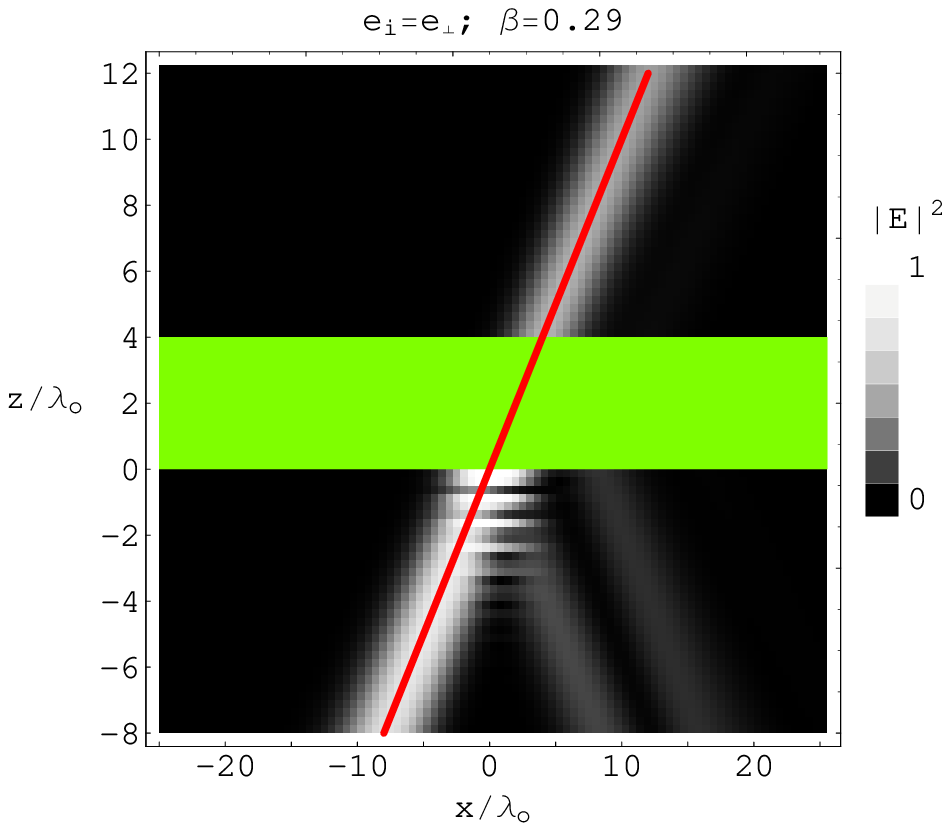,width=2.9in} \\
  \epsfig{file=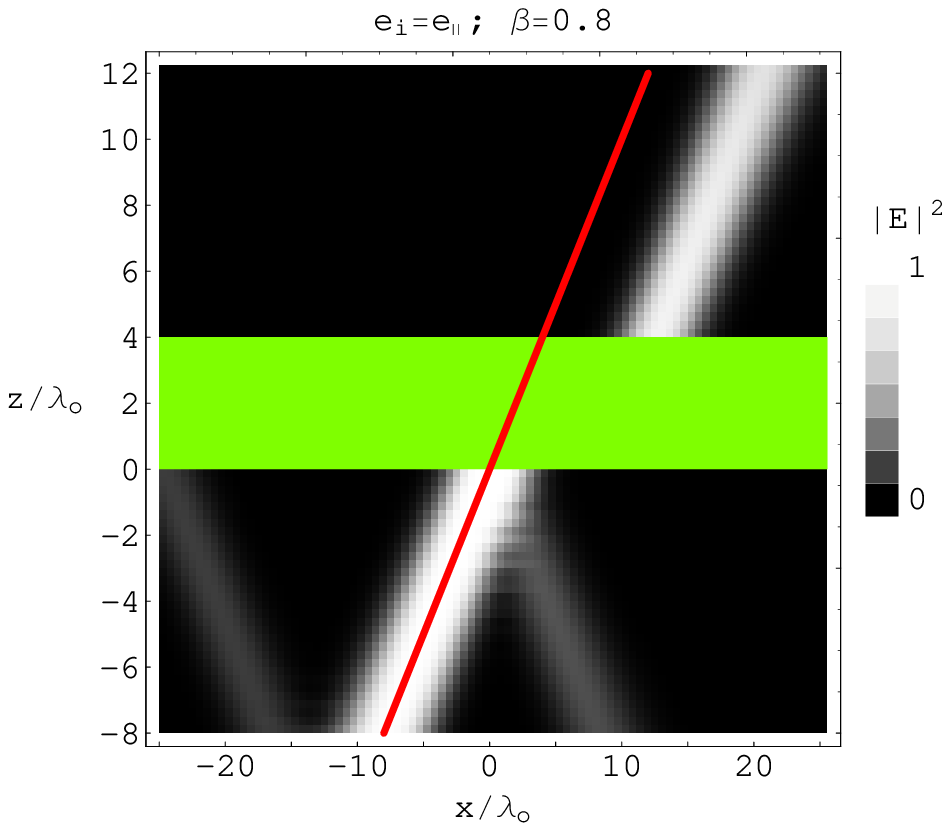,width=2.9in} \hfill \epsfig{file=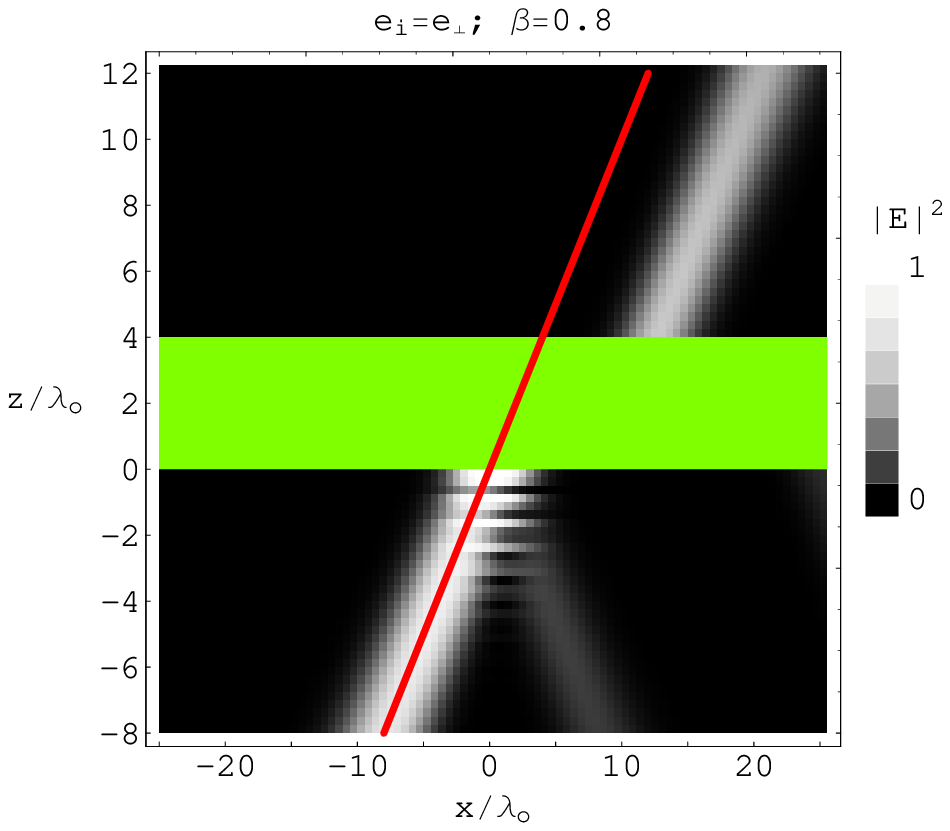,width=2.9in}
 \caption{\label{fig4} Normalized $| \#E |^2$ is mapped in the $xz$ plane for a  2D Gaussian
 beam incident onto a simply moving slab at an angle  $\theta_i =
 45^\circ$. The relative speed of the slab is: $\beta = -0.15$
 (top); $\beta = 0.29$
 (middle); and $\beta = 0.8$
 (bottom). The electric field phasor of the incident beam is
 polarized parallel (left) and perpendicular (right)  to the plane of
 incidence. The  red line indicates the mean beam position in
 the absence of the moving slab.
  }
\end{figure}

In view of Figure~\ref{fig2}, we fixed the
  mean angle of incidence of the beam at $\theta_i = 45^\circ$ and
  explored
  the behaviour of the transmitted beam for $\beta < 0.29$, $\beta = 0.29$ and
  $\beta > 0.29$.
The energy density of the beam in both half--spaces, as measured
by
\begin{equation}
 |\#E |^2 = \left\{
 \begin{array}{lcr}
|\#E_i   + \#E_r |^2 & \mbox{for} & z \leq 0 \vspace{4pt} \\
|\#E_t |^2 & \mbox{for} & z \geq L
\end{array}
\right. ,
\end{equation}
is mapped  for $z / \lambdao \in(-8, 12)$ and $ x / \lambdao \in
(-25, 25)$ in Figure~\ref{fig4} with the slab thickness $L = 4
\lambdao$.
 The restriction $ \vartheta \in \left[-1,1\right]$ was
imposed to exclude evanescence.  A beam waist of $\wo = 1.75
\lambdao$ was selected for all calculations.  We considered $\beta
\in \lec -0.15, 0.29, 0.8 \ric$ for both $\#e_i = \#e_\parallel$ and
$\#e_i = \#e_\perp$.

Regardless
of the polarization state,
 the transmitted beam does not undergo a lateral shift (relative to
the incident beam) when $\beta = 0.29$. However, the transmitted
beam is laterally shifted in the direction of $- \hat{\#x}$ when
$\beta < 0.29$ and in the direction of $+ \hat{\#x}$  when $\beta >
0.29$. The energy densities of the reflected  and transmitted beams
are sensitive to $\beta$ and the polarization state of the incident
beam.

\begin{figure}[!ht]
\centering \psfull \epsfig{file=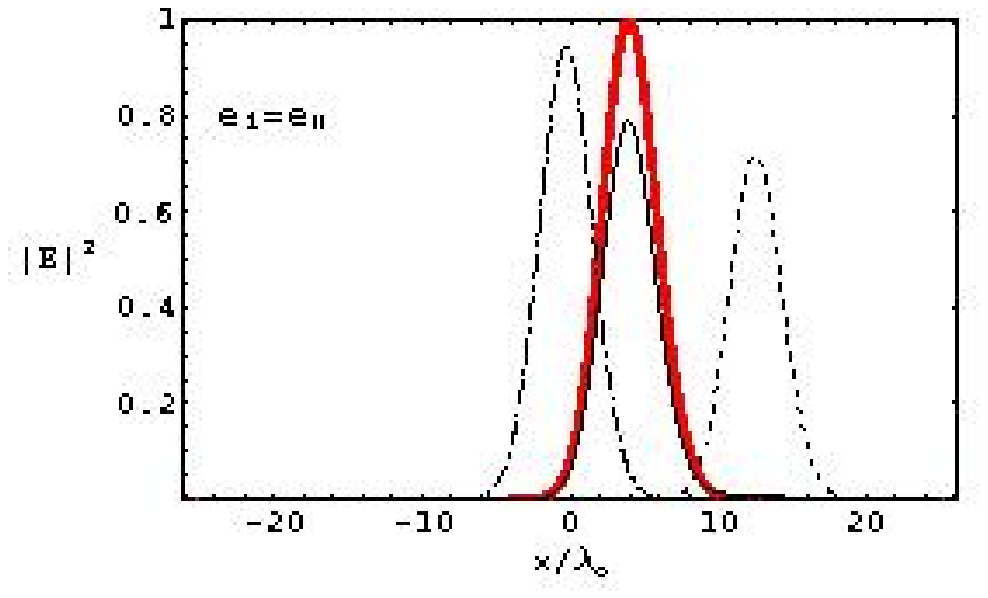,width=3.0in} \hfill
\epsfig{file=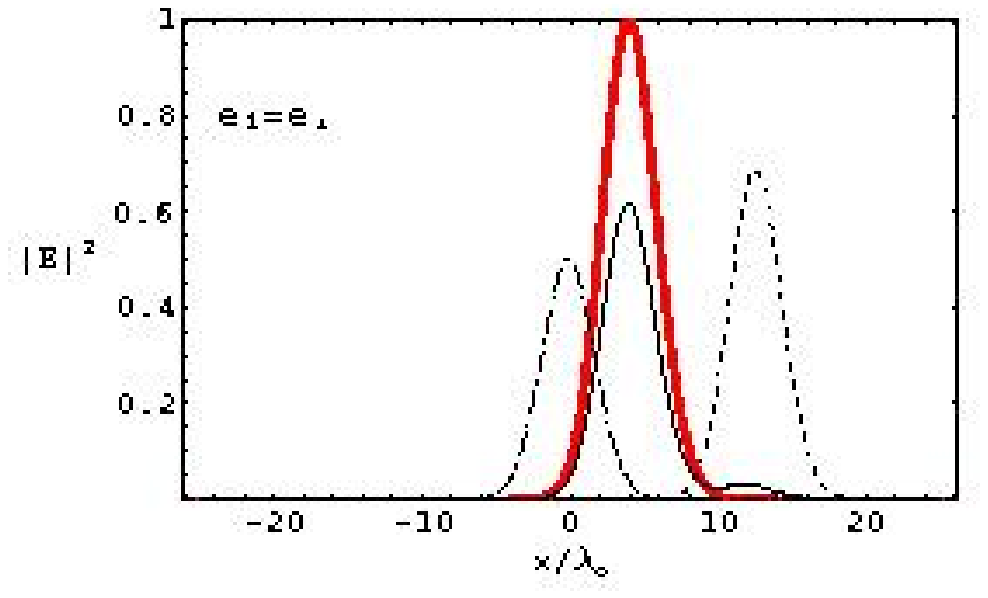,width=3.0in}
 \caption{\label{fig5}
 Normalized $| \#E |^2$ at $z = 4  \lambdao$ for $\beta = -0.15$
 (broken dashed curve); $\beta = 0.29$
 (solid dark curve); and $\beta = 0.8$
 (dashed curve). The solid red curve represents the normalized $| \#E
 |^2$ in the absence of the moving slab. The electric field phasor of the incident beam is
 polarized parallel (left) and perpendicular (right)  relative to the plane of
 incidence.
  }
\end{figure}

A more quantitative representation of the transmitted beam is
provided in Figure~\ref{fig5}, wherein $| \#E |^2$ is plotted
against $x$ for $\beta \in \lec -0.15, 0.29, 0.8 \ric$ at $z = 4
\lambdao$.
 For comparison, $| \#E |^2$ for the beam in the absence
of the moving slab is also plotted. It is clear that the beam
position for $\beta = 0.29$ coincides with the beam position in the
absence of the moving slab. At $\beta = 0.29$,  the peak energy
density of the transmitted beam for the case of parallel
polarization is approximately 11$\%$ less than it would be if the
slab were absent; the corresponding figure for perpendicular
polarization is $38\%$.

\begin{figure}[!ht]
\centering \psfull \epsfig{file=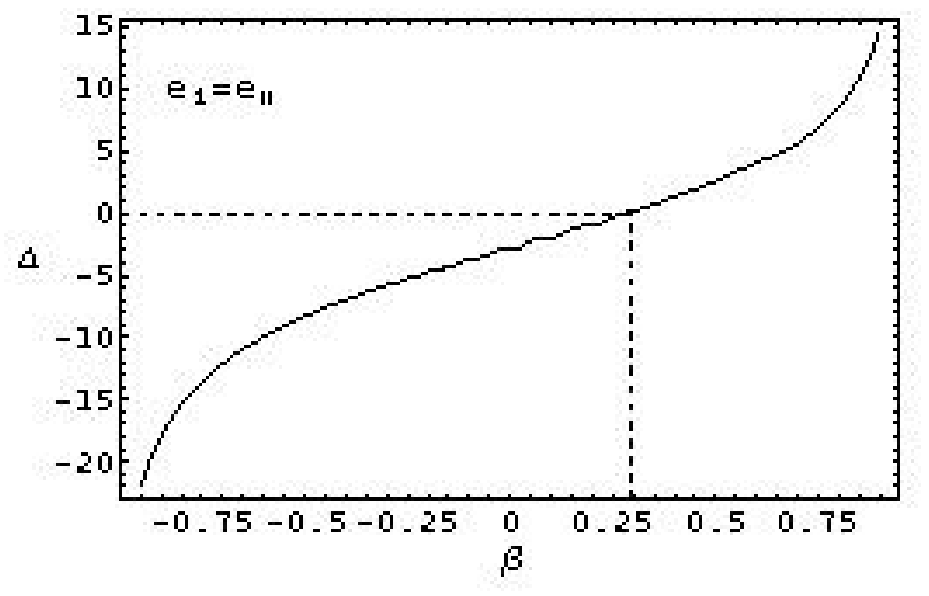,width=3.0in}\hfill
\epsfig{file=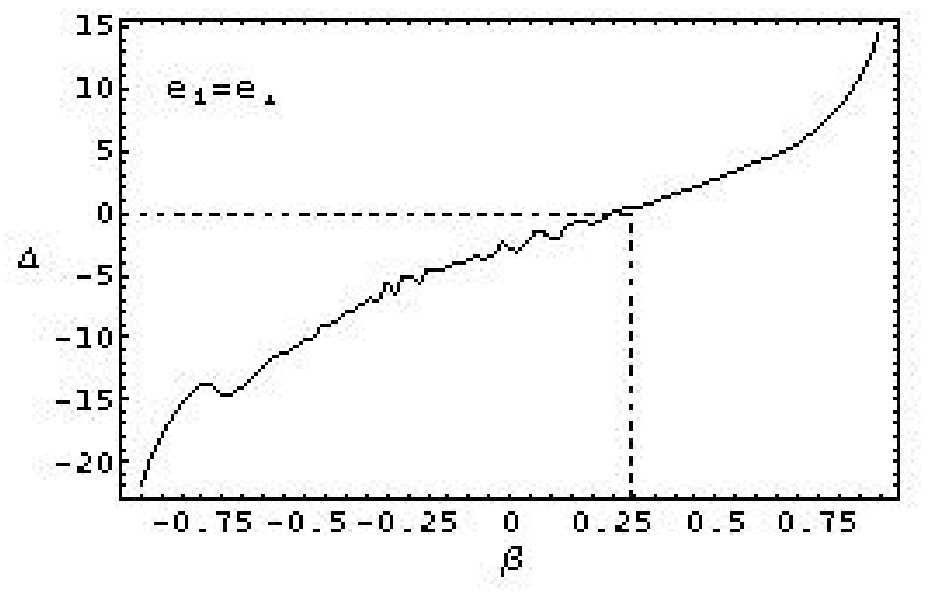,width=3.0in}
 \caption{\label{fig6} The median beam shift $\Delta$
 at $z = 4 \lambdao$, plotted against $\beta$. The electric field phasor of the incident beam is
 polarized parallel (left) and perpendicular (right) relative to the plane of
 incidence. Zero median beam shift at $\beta = 0.29$ is indicated by dashed
 lines.
  }
\end{figure}

The median shift of transmitted beam in relation to the
incident beam is defined as
\begin{eqnarray}
\Delta &=& \le \int^\infty_{-\infty} x | \#E_t (x,L)|^2 \; dx \ri
\le \int^\infty_{-\infty}  | \#E_t (x,L) |^2 \; dx \ri^{-1} \nonumber \\
&& - \le \int^\infty_{-\infty} x | \#E_i (x,L) |^2 \; dx \ri \le
\int^\infty_{-\infty}  | \#E_i (x,L)|^2 \; dx \ri^{-1}.
\end{eqnarray}
For both parallel and perpendicular polarizations, $\Delta$ is
plotted against $\beta \in (-1,1)$ in Figure~\ref{fig6}. Thus,   regardless of the polarization state, the beam can be shifted
laterally along $\pm \hat{\#x}$ by means of uniform motion. In
particular, the zero beam shift at $\beta = 0.29$ is further
confirmed in Figure~\ref{fig6}.

\section{Concluding remarks}

Our numerical investigations show that a 2D beam can pass obliquely
through a uniformly moving slab without undergoing a lateral shift
in its position. At a fixed angle of beam incidence, this effect
occurs only for a unique translational slab velocity. However,
extrapolating from Figure~\ref{fig2},  for every angle of beam
incidence a slab velocity can be found at which the beam undergoes
no lateral shift. Furthermore, for a fixed angle of incidence, a pulsed
beam will undergo zero lateral deflection,
provided that the constitutive parameters do not vary with angular frequency
in the pulse spectrum.

The degree of concealment achieved by uniform motion is  not 100$\%$
due to reflections but, in the particular case of the  example
considered in \S\ref{Beam_section}, almost 90$\%$ of the peak energy
density of the beam can  be transmitted without deflection.

\vspace{10mm}

\noindent{\bf Acknowledgement:}  TGM is supported by a \emph{Royal
Society of Edinburgh/Scottish Executive Support Research
Fellowship}.

\section*{Appendix}

The reflection coefficients $r_{\parallel, \perp}$ and transmission
coefficients $t_{\parallel, \perp}$ are straightforwardly calculated
by solving the reflection--transmission problem as a boundary value
problem. We outline the procedure here, further details being
available elsewhere (Lakhtakia \& Messier 2005).

Consider the plane wave with electric and magnetic field phasors
\begin{equation}
\left.
\begin{array}{l}
\#E(x,z) = \tilde{\#e}(z,\theta) \, \exp \le i \ko x \sin \theta \ri
\vspace{4pt}
\\
\#H(x,z) = \tilde{\#h}(z,\theta) \, \exp \le i \ko x \sin \theta \ri
\end{array} \right\} \l{eh}
\end{equation}
propagating in the $xz$ plane.  As in \S\ref{Beam_section}, a moving
slab described by the Minkowski constitutive relations \r{CRs}
occupies the region between $z= 0$ and $z=L$; elsewhere there is
vacuum.
 We write
\begin{equation}
\tilde{\#p} (z, \theta) = \tilde{p}_x (z, \theta) \, \hat{\#x} +
\tilde{p}_y (z, \theta) \, \hat{\#y} + \tilde{p}_z (z, \theta) \,
\hat{\#z}, \hspace{20mm} (p = e,h).
\end{equation}
 Substitution of equations  \r{CRs} and \r{eh} into the source--free
Maxwell curl postulates
\begin{eqnarray}
&& \nabla \times \#E(x,z) - i \omega \#B (x,z) = \#0,\\
&& \nabla \times \#H(x,z) + i \omega \#D (x,z) = \#0,
\end{eqnarray}
delivers four  differential equations and two algebraic equations.
The latter two equations are easily solved for $\tilde{e}_z$ and
$\tilde{h}_z$. Thereby, the four differential equations may be
expressed in  matrix form as
\begin{equation}
\frac{\partial}{\partial z} \les \#f (z, \theta) \ris = i \ko \les
\#P(\theta) \ris \les \#f (z, \theta) \ris, \l{MODE}
\end{equation}
where
\begin{equation}
\les \#f (z, \theta) \ris = \les \tilde{e}_x  (z, \theta),
\,\tilde{e}_y  (z, \theta), \,\tilde{h}_x  (z, \theta),
\,\tilde{h}_y (z, \theta) \ris^T
\end{equation}
is a column vector and
\begin{equation}
\#P(\theta) = \les \begin{array}{cccc} 0&0&0& \etao \rho \\
0&0& -\etao &0\\
0& -\eps_r \rho / \etao &0&0\\
\eps_r / \etao & 0 &0&0
\end{array}
\ris
\end{equation}
is a 4$\times$4 matrix with
\begin{equation}
\rho = \alpha - \frac{\le m + \sin \theta \ri^2}{\eps_r \alpha}.
\end{equation}
The solution to \r{MODE} is conveniently expressed as
\begin{equation}
\les \#f (L, \theta) \ris = \les \#M(L, \theta) \ris \les \#f (0,
\theta) \ris, \l{MODE_sol}
\end{equation}
in terms of the transfer matrix
\begin{equation}
\les \#M(L, \theta) \ris = \exp \lec i \ko \les \#P (\theta) \ris \,
L \ric.
\end{equation}

Now we turn to the incident, reflected and transmitted plane waves.
Let the incident plane wave be represented in terms of linear
polarization components as
\begin{equation}
\left.
\begin{array}{l}
\tilde{\#e}_i (z, \theta) = \les a_\perp \, \hat{\#y} +  a_\parallel
\le - \cos \theta \,\hat{\#x} + \sin \theta \,\hat{\#z} \ri \ris
\exp \le i \ko z \cos \theta \ri \vspace{4pt}
\\
\tilde{\#h}_i (z, \theta) = \etao^{-1} \les a_\perp  \le - \cos
\theta \,\hat{\#x} + \sin \theta \,\hat{\#z} \ri  - a_\parallel \,
\hat{\#y}\,
 \ris
\exp \le i \ko z \cos \theta \ri
\end{array}
\right\}, \hspace{10mm} z \leq 0.
\end{equation}
The corresponding reflected and transmitted plane waves are given as
\begin{equation}
\left.
\begin{array}{l}
\tilde{\#e}_r (z, \theta) = \les a_\perp r_\perp \, \hat{\#y} +
a_\parallel r_\parallel \le  \cos \theta  \,\hat{\#x} + \sin \theta
\,\hat{\#z} \ri \ris \exp \le -i \ko z \cos \theta \ri \vspace{4pt}
\\
\tilde{\#h}_r (z, \theta) = \etao^{-1} \les a_\perp r_\perp \le \cos
\theta \,\hat{\#x} + \sin \theta \,\hat{\#z} \ri  - a_\parallel
r_\parallel \, \hat{\#y}\,
 \ris
\exp \le -i \ko z \cos \theta \ri
\end{array}
\right\}, \hspace{10mm} z \leq 0
\end{equation}
and
\begin{equation}
\left.
\begin{array}{l}
\tilde{\#e}_t (z, \theta) = \les a_\perp t_\perp \, \hat{\#y} +
a_\parallel t_\parallel \le - \cos \theta  \,\hat{\#x} + \sin \theta
\,\hat{\#z} \ri \ris \exp \les i \ko (z-L) \cos \theta \ris
\vspace{4pt}
\\
\tilde{\#h}_t (z, \theta) = \etao^{-1} \les a_\perp t_\perp \le -
\cos \theta \,\hat{\#x} + \sin \theta \,\hat{\#z} \ri  - a_\parallel
t_\parallel \, \hat{\#y}\,
 \ris
\exp \les i \ko (z-L) \cos \theta \ris
\end{array}
\right\}, \hspace{10mm} z \geq L,
\end{equation}
respectively. By application of the boundary conditions at $z=0$ and
$z=L$ to the solution \r{MODE_sol}, the reflection and transmission
coefficients are found to be related by the matrix algebraic
equation
\begin{equation}
\les \#K (\theta) \ris \les  t_\perp,  t_\parallel,   0,   0 \ris^T
= \les \#M (L, \theta) \ris \les \#K (\theta) \ris
 \les  1, 1, r_\perp,  r_\parallel  \ris^T,
 \end{equation}
wherein
\begin{equation}
\#K (\theta) = \les \begin{array}{cccc} 0 & - \cos \theta & 0 & \cos
\theta \\
 1 & 0 & 1 & 0 \\
 - \etao^{-1} \cos \theta & 0 & \etao^{-1} \cos \theta & 0 \\
 0 & - \etao^{-1} & 0 & - \etao^{-1}
\end{array}
\ris.
\end{equation}
Thus, after some manipulation,  the reflection and transmission
coefficients emerge as
\begin{eqnarray}
r_\perp &=& \frac{\le  \cos^2 \theta - \eps_r \rho  \ri \sin \le \ko
L \sqrt{\eps_r \rho} \ri}{\le  \cos^2 \theta + \eps_r \rho  \ri \sin
\le  \ko L \sqrt{\eps_r \rho} \ri  + 2 i \sqrt{\eps_r \rho} \, \cos
\le \ko L \sqrt{\eps_r \rho} \ri \, \cos \theta}, \l{r_perp} \vspace{4pt}\\
r_\parallel &=& \frac{\le \rho - \eps_r \cos^2 \theta   \ri \sin \le
\ko L \sqrt{\eps_r \rho} \ri}{\le  \eps_r  \cos^2 \theta +  \rho \ri
\sin \le  \ko L \sqrt{\eps_r \rho} \ri  + 2 i \sqrt{\eps_r \rho} \,
\cos \le \ko L \sqrt{\eps_r \rho} \ri \, \cos \theta}, \vspace{4pt}
\\
t_\perp &=& \frac{- 2 i \sqrt{\eps_r \rho} \, \cos \theta}{\le
\cos^2 \theta + \eps_r \rho \ri \sin \le \ko L \sqrt{\eps_r \rho}
\ri  + 2 i \sqrt{\eps_r \rho} \, \cos
\le \ko L \sqrt{\eps_r \rho} \ri \, \cos \theta}, \vspace{4pt} \\
t_\parallel &=& \frac{2 i \sqrt{\eps_r \rho} \, \cos \theta}{\le
\eps_r  \cos^2 \theta +  \rho \ri \sin \le  \ko L \sqrt{\eps_r \rho}
\ri  + 2 i \sqrt{\eps_r \rho} \, \cos \le \ko L \sqrt{\eps_r \rho}
\ri \, \cos \theta}. \l{t_perp}
\end{eqnarray}

\newpage

\vspace{15mm}

\vspace{5mm}

\newpage

\newpage

\vspace{10mm}

\begin{thebibliography}{99}
\nonumber

\bibitem{Alu}
Al\`{u}, A. \& Engheta, N. 2005 Achieving transparency with
plasmonic and metamaterial coatings. \emph{Phys. Rev. E} {\bf 72},
016623. Erratum: 2006 {\bf 73}; 019906(E).

\bibitem{Chen}
Chen, H.C. 1983 {\em Theory of electromagnetic waves\/}. New York,
NY, USA: McGraw--Hill.

\bibitem{Fedotov}
Fedotov, V.A., Mladyonov, P.L., Prosvirnin, S.L \& Zheludev, N.I.
2005 Planar electromagnetic metamaterial with a fish scale
structure. \emph{Phys. Rev. E} {\bf 72}, 056613.


\bibitem{Haus}
Haus, H.A. 1984 {\em Waves and fields in optoelectronics\/}.
Englewood Cliffs, NJ, USA: Prentice--Hall.

\bibitem{Optik}
 Lakhtakia, A. \&   McCall, M.W. 2004 Counterposed phase velocity and
energy--transport velocity vectors in a dielectric--magnetic
uniaxial medium. \emph{Optik} {\bf 115},  28--30.


\bibitem{STF_book}
Lakhtakia A. \& Messier R. 2005 {\em Sculptured thin films\/}.
Bellingham, WA, USA: SPIE Press.

\bibitem{Leonhardt}
Leonhardt, U. 2006  Optical conformal mapping. \emph{Science} {\bf
312}, 1777--1780.



\bibitem{ML04}
Mackay, T.G. \&   Lakhtakia, A. 2004  Negative phase velocity in a
uniformly moving, homogeneous, isotropic, dielectric--magnetic
medium. \emph{J. Phys. A: Math. Gen.} {\bf 37}, 5697--5711.



\bibitem{ML06a}
Mackay, T.G. \&   Lakhtakia, A. 2006a  On electromagnetics of an
isotropic chiral medium moving at constant velocity \emph{Proc. R.
Soc. Lond. A} (to appear).


\bibitem{ML_counterposition}
Mackay, T.G. \&  Lakhtakia, A. 2006b Counterposition and negative
refraction due to uniform motion. {\sf
http://arxiv.org/abs/physics/0610039 }

\bibitem{Milton_cloak}
Milton, G.W. \& Nicorovici, N--A. P. 2006 On the cloaking effects
associated with anomalous localized resonance. \emph{Proc. R. Soc. Lond.
A} {\bf 462}, 3027--3059.

\bibitem{Pendry_cloak}
Pendry, J.B., Schurig, D. \& Smith, D.R. 2006 Controlling
electromagnetic fields. \emph{Science} {\bf 312}, 1780--1782.

\bibitem{Wolf}
Wolf, E. \& Habashy, T. 1993 Invisible bodies and uniqueness of the
inverse scattering problem. \emph{J. Mod. Optics} {\bf 40},
785--792.

\end{thebibliography}
\end{document}